\documentclass[prl,aps,reprint,groupedaddress,showkeys]{revtex4-1}

\usepackage[utf8]{inputenc}

\usepackage{fullpage}
\usepackage[T1]{fontenc}
\usepackage{verbatim}
\usepackage{amssymb,amsmath}
\usepackage{comment,xcolor}
\usepackage[colorlinks,linktocpage,linkcolor=blue,citecolor=red,urlcolor=red]{hyperref}

\def\be{\begin{equation}} 
\def\ee{\end{equation}}

\allowdisplaybreaks

\begin{document}

\title{Warped G$_2$-throats in IIA and uplift dSillusions}

\author{Fotis Farakos$^{a}$}\author{George Tringas$^b$} \author{Thomas Van Riet$^c$}

\affiliation{}
\affiliation{$^a$ Physics Division, National Technical University of Athens
15780 Zografou Campus, Athens, Greece} 
\affiliation{$^b$ Dep.~of Physics, Lehigh University, 16 Memorial Drive East, Bethlehem, PA 18018, USA} 
\affiliation{$^c$ Inst.~for Theoretical Physics, KU Leuven, Celestijnenlaan 200D, B-3001 Leuven, Belgium}
\email{fotis.farakos@gmail.com,\\ georgios.tringas@lehigh.edu,\\ thomas.vanriet@kuleuven.be}

\begin{abstract}

Flux compactifications of IIA supergravity on orientifolded G$_2$-manifolds have been argued to allow for classical Minkowski$_3$ vacua with moduli and scale-separated AdS$_3$ vacua with full moduli stabilisation. To further uplift these vacua to meta-stable dS$_3$ vacua using anti-D2 branes, warped throats are desirable. We study the flux-stabilisation of local ``CGLP-type'' throats in compact G$_2$ spaces, and discuss consistency constraints on anti-brane uplifting. Despite the classical AdS$_3$ vacua to be free of tachyons, we find that uplifting from anti-branes down warped throats is forbidden. If instead we rely on hypothetical AdS$_3$ vacua that arise from quantum corrections to the classical Minkowski$_3$ vacua, we find that (similarly to the 4d analogues) consistency constraints point in opposite directions. However, there is potentially an advantage over 4d when it comes to concrete fine-tuning freedom of numbers, such as tadpole constraints.
\end{abstract}

\maketitle

\section{1. \NoCaseChange{Introduction}}

The construction of semi-realistic vacua within computable corners of string theory remains challenging when we require moduli stabilisation, smallness of the vacuum energy (\emph{aka} scale separation), and especially when we want the vacuum energy to be positive. 
For AdS vacua there are concrete candidates for achieving moduli stabilization and scale separation but none of them are considered fully achieved yet as some approximations remain in the computations; see \cite{Coudarchet:2023mfs, Tringas:2025uyg} for recent reviews. Instead, for solutions with rolling scalar fields, as opposed to vacua, it has been pointed out that scale separation is fairly generic \cite{Andriot:2025cyi}. 

Finding vacua with positive vacuum energy is notoriously difficult \cite{Danielsson:2018ztv, Cicoli:2018kdo} and as of today not a single, sufficiently explicit, proposal has survived scrutiny \cite{VanRiet:2023pnx}. 
The main issue for dS solutions is the absence of parametric control. This can be argued at the technical level \cite{Junghans:2018gdb, Banlaki:2018ayh}, and understood heuristically using arguments based on entropy \cite{Ooguri:2018wrx} or the species scale \cite{Hebecker:2018vxz}. Finally, there is also a problem of perturbative stability and tachyons seem hard to avoid \cite{Danielsson:2012et, Andriot:2021rdy}.  
Much of this trouble can be understood from various Swampland principles \cite{Vafa:2005ui, Palti:2019pca}, which leads us to investigate the situation in 3d instead of 4d since 3d is a critical dimension for the Swampland program.  Regardless of phenomenology, achieving a controlled dS$_3$ vacuum of string theory would be a big leap forward and of relevance to putative dS holography.

In 3d there exists a class of classical ``no-scale''  Minkowski vacua with flat directions \cite{Blaback:2010sj, Farakos:2020phe} and classical AdS vacua with moduli stabilization, arbitrary separation of scales, and minimal or no supersymmetry \cite{Farakos:2020phe} (\cite{VanHemelryck:2022ynr, Emelin:2021gzx, Farakos:2023nms,  Arboleya:2024vnp, VanHemelryck:2025qok,Farakos:2025bwf}). These 3d vacua are free of tachyons, whereas their 4d counterparts \cite{DeWolfe:2005uu} always have tachyons, unless one considers Calabi-Yau's without complex structure deformations \cite{Apers:2022tfm}. 
Concerning 3d dS constructions, we are only aware of \cite{Dong:2010pm} and a previous paper of ours \cite{Farakos:2020idt} looking for dS$_3$ solutions ``in the neighborhood'' of the AdS$_3$ vacua in the hope that moduli stabilisation, scale separation and a parametric classical limit is not lost too much. Yet, the results of \cite{Farakos:2020idt} were mixed: although the closed-string tachyons (ubiquitous in 4d) seem absent, it was claimed that instead open-string tachyons would appear. Secondly, it was found that obeying tadpole cancellation and flux quantization at the same time seemed impossible unless one relies on models with warped throats. In this paper we point out that the tachyon problem was due to a typo \cite{Cassandra} and most of the paper is devoted to demonstrating how manifolds with the required properties such as G$_2$-holonomy and warped throats seem reasonable and how uplift energies depend on the conifold modulus, the string coupling and the volume.  Yet, we will find that the classical AdS$_3$-vacua  can not be used to uplift to dS$_3$ vacua by means of anti-branes down warped throats. Hence, if warped throats with anti-branes are to be used in 3d, one again has to resort to quantum corrections to no-scale Minkowski vacua as in 4d. Despite the suggestion by Kachru-Kallosh-Linde-Trivedi (KKLT) \cite{Kachru:2003aw} it has now been argued that a successful uplift to dS$_4$ is obstructed  \cite{Carta:2019rhx, Gao:2020xqh}. 
We argue that the difficulties in 3d are similar but that perhaps the tadpole numbers of G$_2$ manifolds might be larger than for CY$_3$.  

Our arguments for the existence of G$_2$ spaces with stabilised throats mirrors the Giddings-Kachru-Polchinski (GKP) embedding \cite{Giddings:2001yu} of a Klebanov-Strassler (KS) throat \cite{Klebanov:2000hb} inside a 
compact Calabi-Yau (CY).

\section{2. \NoCaseChange{Classical Minkowski$_3$ and AdS$_3$}}

KKLT \cite{Kachru:2003aw} suggested that meta-stable dS can be obtained from SUSY AdS \emph{with positive masses} through an uplift with a redshifted anti-brane. They further argued that the quantum corrections to the no-scale Minkowski vacua obtained from 3-form fluxes in IIB string theory \cite{Giddings:2001yu}, lead to scale-separated AdS$_4$ vacua with all moduli stabilised and without tachyons. This way they form a possible starting point for anti-brane uplifts.

We follow this suggestion in 3d but with the difference that our scale-separated AdS$_3$ vacua without tachyons can already be generated at the classical level. We will also briefly discuss non-classical stabilisation of the moduli present in the classical no-scale solutions in 3d. All of these vacua can be found in the context of (massive) IIA on G$_2$ manifolds with fluxes and O2/O6 planes as first pioneered in \cite{Farakos:2020phe}.   

The 10d metric in Einstein frame has the form \cite{Blaback:2010sj}
\be 
\label{warped-metric}
 ds^2 = \frac14 e^{2A} e^{2 \alpha v} d\tilde s^2_3 + e^{-\frac{6}{5}A} e^{2 \beta v} d\tilde s^2_{7} \,, 
\ee 
with $e^A$ the a warp-factor and $v$ is the volume modulus, and $d\tilde s^2_{7}$ is the unit-volume metric on the compact 7d space. 3d Einstein frame and canonical normalisation requires  $\alpha = - 7 \beta$ and $\alpha^2 = 7/16$.  In minimal 3d supergravity the scalar potential $V$ can be written in terms of a real superpotential $P$ as:
\begin{equation}
\label{Vsusy}
V = 2 G^{ij}\partial_iP\partial_jP -4P^2\,,   
\end{equation}
where $G_{ij}$ is determined by the scalar kinetic term $\mathcal{L}\supset-\tfrac{1}{2}G_{ij}\partial\phi^i\partial \phi^j$ from dimensional reduction. 
For massive IIA supergravity compactified on G$_2$ orientifolds with $F_4$ and $H_3$ fluxes, the superpotential has the form \cite{Farakos:2020phe}
\be
\label{P-SUSY}
P=\frac{1}{8}e^{-11\beta\upsilon-\frac{\phi}{2}}\left(-f e^{\frac{3}{4}\phi}+e^{\beta\upsilon}h+e^{4\beta\upsilon+\frac{7}{4}\phi}m\right) \,.
\ee 
All other scalar fields, describing model-dependent fluctuations of the specific G$_2$ spaces and the form fields in 10d are hidden in the symbols $h$ and $f$. Within our two-scalar truncation we have $G_{ij}=\text{diag}[1/2,1/2]$. The reason for this shorthand notation and truncation is that $\overline{D2}$-branes down throats decouple from these bulk fields. As a consequence the positions of these ``non-universal'' scalars are identical to these of the SUSY AdS$_3$ vacua of \cite{Farakos:2020phe}. What counts for this work is that these scalars are stabilised with all positive masses \cite{Farakos:2020phe, Farakos:2020idt}. This is crucial for dS model building since scalars with negative $m^2$ would make the de Sitter uplift unstable. 

The symbol $f$ comprises two kinds of $F_4$ fluxes, dubbed $F_{4A}$ and $F_{4B}$ in \cite{Farakos:2020phe}. The  $F_{4A}$-flux obeys $F_{4A} \wedge H_3 =0$ and as such are not bounded by tadpoles, whereas $F_{4B} \wedge H_3 \ne 0$ and so is bounded by the RR tadpoles. Parametric scale separation requires the limit of large $F_{4A}$ and hence large $f$.

We now recall the (SUSY) vacua in this model  \cite{Farakos:2020idt} from the point of the view of the universal scalars $\phi,v$.  Without Romans mass we can satisfy the F-term conditions $\partial_{\phi,v}P=0$ as long as $f h <0$ to find $P=0$ on-shell, realizing Minkowski SUSY no-scale. If instead $hf>0$, then we have no-scale with broken supersymmetry.  With Romans mass $m$ consistency with the O6 tadpoles, requires $h m > 0$. Extremization of the superpotential for $m>0$ and $h>0$ gives the consistency condition $f>0$. The vacuum energy $V$, string coupling $g_s$ and volume are then given by 
\begin{align}
\label{GS}
&V = - \frac{256 h^6 m^4}{729 {f}^8}\,,\quad g_s =e^{\phi}= \frac{2^{7/4} h}{3 (f^3 m)^{1/4}} \,,\\ 
&\text{vol}(X_7) =e^{7\beta\upsilon}= \left(\frac{f}{2 m}\right)^{7/4} \,, 
\label{VOL}
\end{align}
in string units. Therefore, large $f$ leads to parametric control and  scale separation. As explained in detail in \cite{Farakos:2020phe, Farakos:2020idt} the O2/O6 tadpole cancellation works such $m$ remains order one and $f$ can be taken arbitrary large. The moduli masses, within this two-scalar truncation, are positive
\be
\frac{V''}{|V|} = \frac{4}{49} \left( 149 \pm 13 \sqrt{67} \right) > 0 \,, 
\ee
and they remain positive when the mixing with all other fields is considered in concrete models. 

We are interested in realising a concrete uplift to dS$_3$ by breaking supersymmetry with an $\overline{D2}$-brane, analogues to the original proposals by Kachru-Pearson-Verlinde \cite{Kachru:2002gs} and KKLT \cite{Kachru:2003aw}. Those proposals rely on $\overline{D3}$-branes which are dynamically attracted towards the tip of a KS-like throat where its energy is redshifted exponentially. Whereas the KPV scenario computes the perturbative stability explicitly for the non-compact KS throat, the KKLT scenario relies on the GKP mechanism \cite{Giddings:2001yu} to embed this throat into a compact CY-space. We will parallel these steps and embed  Cvetic-Gibbons-Liu-Pope  (CGLP)-type  throats~\cite{Cvetic:2001bw,Cvetic:2001ih,Cvetic:2001kp} into compact G$_2$-spaces and study the stabilisation of the conifold modulus. An exponentially small conifold modulus will relate to exponentially large warping down such a throat. 

\section{3. \NoCaseChange{Conifolds and throats}}

Our uplift scenario requires the existence of local CGLP-type throats with finite 4-cycles at the tip, stabilised at exponentially small values of the conifold modulus, all inside a compact G$_2$-space. What follows provides a proof of principle through an example, but our arguments for meta-stable dS$_3$ should be independent of any details. The interested reader is referred to \cite{Acharya:2020vmg} for recent developments on G$_2$-conifolds. 

Our example relies on a constructing \emph{a barely G$_2$-cone} using the deformed conifold of KS.  The seven-dimensional metric 
\be
d\tilde s^2_{7} = dY^2 + ds^2_{CY} \,, 
\ee
describes a product, $CY_3\times S^1$ with $Y$ the circle coordinate. Both of these spaces are unit-volume as the overall volume has been extracted; see \eqref{warped-metric}. To turn the holonomy into G$_2$ and not $SU(3)$ one has to mod-out by a further $Z_2$ involution $\hat{\sigma}$ \cite{Harvey:1999as,Kachru:2001je}, which combines an inversion of $\sigma: Y \mapsto -Y$, with an anti-holomorphic $Z_2$ involution $\mathcal{I}: z_i \mapsto \overline{z}_i$. We therefore have $\hat{\sigma} = \sigma \times \mathcal{I}$, and the (unresolved) G$_2$-space $X$ is 
\be\label{X}
X = \frac{Y \times S^1}{\hat \sigma} \,. 
\ee
The CY$_3$-space has the $(1,1)$ K\"ahler 2-form $J_2$ and the holomorphic $(3,0)$ 3-form $\Omega_3$.  Using those we can define the associative G$_2$ 3-form 
\be
\Phi = dY \wedge J_2 + \text{Re}\,\Omega_3 \,, 
\ee
while the associative 4-form is  
\be
\star \Phi = \frac12 J_2 \wedge J_2 - d Y \wedge \text{Im}\,\Omega_3 \,. 
\ee
Under $z_i \mapsto \overline{z}_i$, the holomorphic 3-form  
behaves as $\mathcal{I}: \Omega_3 \mapsto \overline{\Omega}_3$ and $\mathcal{I}: J_2 \to - J_2$ and so $\Phi$ is invariant. The deformed conifold is described by 
\be
\label{DCz}
\sum_{i=1}^4 z_i^2 = - s \,,
\ee 
where $s$ is a real number, called the conifold modulus, and the minus sign arises due to a rotation of the standard deformed conifold coordinates in our chosen frame. Note that the involution $\mathcal{I}: z_i \mapsto \overline{z}_i$ leaves the definition of the deformed CY conifold \eqref{DCz} invariant. For $s$ positive the involution does not have any fixed points and so we get a barely G$_2$-space \cite{Harvey:1999as} with a finite 3-cycle (\emph{aka} A-cycle) at the tip of the throat controlled by the value of $s$. The dual 3-cycle (\emph{aka} B-cycle) extends along the warped direction. In a non-compact setting, the stabilisation of $s$ with 3-form flux was found explicitly by solving for the 10d (SUSY) equations \cite{Klebanov:2000hb} and in a compact setting this can be found by solving the $F$-term equation with respect to the conifold modulus \cite{Giddings:2001yu}. We will now repeat the latter procedure for our set-up. From the properties of the deformed conifold in a compact space (and taking into account that we have rotated coordinates) we have \cite{Giddings:2001yu} 
\be
\int_A \text{Im} \, \Omega_3 = s \ , \ 
\int_B \text{Re} \, \Omega_3 =  c + \frac{s}{2 \pi} \log s + \dots 
\ee
with $c$ a number that does not concern us. These properties lead to 
\be
\int_B \Phi = c + \frac{s}{2 \pi} \log s + \dots 
\ , \ 
\int_{A\wedge dY} \star \Phi =  - s + \dots  
\ee
hence $s$ now controls the size of the 4-cycle $A\wedge dY$. Here we assume all the other moduli, but $s$, to be stabilized; with $s$ controlling the size of the 4-cycle $A\wedge dY$, with the topology of an $S^4$. 

To evaluate the value of the $s$ modulus we will search for the extremum of the superpotential induced by the flux. The Romans mass does not impact the stabilization of $s$ as much as the $F_4$ and $H_3$ flux; therefore, we keep only the two latter for our purposes here. The relevant part of the superpotential for our purposes has the form 
\be
P = \frac18 \frac{1}{\text{vol}(X_7)^2} \int_7 \left( \star \Phi \wedge H_3 e^{-\frac{\phi}{2}} + \Phi \wedge F_4 e^{\frac{\phi}{4}} \right) + \dots \,, 
\ee
and using the flux choices 
\be
\int_{A\wedge dy} F_4 =  M \quad , \quad \int_B H_3 = K \,, 
\ee
it reduces to 
\be
\label{P-roughly}
P = \frac{\left(  
- g_s^{-1/2} K s 
+ g_s^{1/4}  M  \left[ c + \frac{s}{2 \pi} \log s  \right] \right)}{8 \left({\cal V}_0 - \left(c + \frac{s}{2 \pi} \log s \right) s\right)^2}  + \dots \,. 
\ee
The full G$_2$-space volume is computed by $\int_7 \Phi \wedge \star \Phi$ and we estimated this integral to be ${\cal V}_0$ plus the parametrically small contribution from the $s$ part;  $\text{vol}(X_7) \simeq {\cal V}_0 - (c + \frac{s}{2 \pi} \log s ) s$. Then, assuming $s \ll 1$ we have the extremization condition 
\be
\label{s-eq}
\frac{\partial P}{\partial s} = 0 
\quad \to \quad  
s \sim e^{- \frac{2 \pi K}{ M g_s^{3/4}}} \,. 
\ee
We find an exponentially suppression. Note that the $M$ corresponds to a $F_{4B}$ flux in the notation of \cite{Farakos:2020phe}, and therefore is bounded by RR-tadpoles. 

Next, we discuss the relation between the warp factor, the conifold modulus and the bulk moduli in 10d Einstein frame since these details matter for the uplift. Near the tip, the conifold region behaves like the CGLP-solutions with a finite 4-cycle volume \cite{Cvetic:2001ma,Herzog:2002ss}
\be
\label{CGLP-tip} 
e^{-\frac{12}{5}A}  \Big{|}_{CGLP-tip}  
\sim \left( M^{-\frac{5}{8}} \times e^{- \frac{5}{32}\phi_0} \right)^{-\frac{12}{5}} \,,
\ee
where $\phi_0$ is the dilaton value at the bulk region and $M$ is again the $F_4$ flux quanta on the finite 4-cycle at the tip. For all these solutions the dilaton behaves as $e^{\phi} = e^{\phi_0} e^{-\frac{4}{5}A}$. To estimate the dependence of the warp factor on the tip with respect to the bulk volume modulus we should match the CGLP-solution with the warped compactification at the throat. The size of the 4-cycle is given by  $s \times e^{4 \beta v}$. This means that the actual size of the 4-cycle taking into account also the warping is in Einstein frame \eqref{warped-metric} given by 
\be
s \times e^{4 \beta v} \times e^{-\frac{12}{5}A} \Big{|}_{tip} \,.
\ee
Since the behavior at the tip is controlled by the local effects, this should match the size of the 4-cycle for the CGLP-solution in Einstein frame \eqref{CGLP-tip}: 
\be
s \times e^{4 \beta v} \times e^{-\frac{12}{5}A} \Big{|}_{tip} 
\sim e^{-\frac{12}{5}A}  \Big{|}_{CGLP-tip}  \,, 
\ee
which gives 
\be
\label{warp-tot} 
e^A \Big{|}_{tip} \sim s^{\frac{5}{12}} \times M^{-\frac{5}{8}} \times e^{- \frac{5}{32}\phi_0} \times e^{\frac{5}{3}\beta v} \,. 
\ee
We conclude that the warping at the tip depends on the bulk values of the  dilaton and volume modulus. 

\section{4. \NoCaseChange{Anti-D2 uplift}}

Now we turn to probe $\overline{D2}$-branes and first demonstrate they are attracted to the tip.  For this we need the $C_3$ profile on the external space, 
which is already given by the CGLP solution and reads \cite{Cvetic:2001ma,Herzog:2002ss}  
\be
C_3^{ext.} =  e^{\frac{16}{5}A - \frac14 \phi_0} \tilde \epsilon_3 \,. 
\ee 
The $\overline{D2}$-brane action on the CGLP background then becomes: 
\be
\begin{aligned}
S_{D2/\overline{D2}} 
= & - \mu_2  e^{3A -\frac14 \phi }  \int_3 \sqrt{- \tilde g_3} \pm \mu_2  \int_3 C_3 
\\ 
= & \mu_2 \sqrt{- \tilde g_3} \, e^{\frac{16}{5}A - \frac14 \phi_0} \, (- 1 \pm 1 ) \,, 
\end{aligned}
\ee
where the plus sign is for a $D2$ whereas the minus for a $\overline{D2}$. The action vanishes for a $D2$-brane and doubles for a $\overline{D2}$. We conclude that the $\overline{D2}$ is dragged to the bottom of the warped throat where its tension redshifts. Crucially, we have assumed that the Romans mass, needed to break the no-scale solution into an AdS compactification, does not alter this picture qualitatively. This is similar to the assumption that in the KKLT scenario the gaugino condensate only affects the UV and not the IR of a throat.

We now compute the moduli dependence of the uplift term by computing the $\overline{D2}$-action:
\begin{align}
S_{\overline{D2}} 
& \simeq - 2 e^{3 \alpha v} e^{3A_\text{Tip}} (2 \pi)^5 N_{\overline{D2}} e^{-\frac{\phi}{4}} \int_3 \sqrt{-\tilde g_3} \,, 
\end{align}
with $\tilde g_{\mu\nu}$ the metric of the 3d space in 3d Einstein frame. The dependence of the warp factor (at the tip) on the bulk moduli and conifold modulus was established in equation \eqref{CGLP-tip}.  We also need to take into account how the dilaton behaves near the tip: 
\be
e^{-\phi/4} \Big{|}_\text{Tip}  \to s^{\frac{1}{12}}  \times M^{-\frac{1}{8}} \times e^{- \frac{9 \phi_0}{32}} \times e^{\frac{ \beta v}{3}} \,. 
\ee
As a result, the effective tension of the $\overline{D2}$ red-shifts and receives an additional dependence on the volume and a different dependence on the bulk dilaton, taking the form 
\be\label{antiD2warping}
S_{\overline{D2}} 
\sim - (2 \pi)^5 N_{\overline{D2}} M^{-2} e^{3 \alpha v+ \frac{16}{3} \beta v} s^{\frac{4}{3}} \ e^{-\frac{3\phi_0}{4}} \int_3 \sqrt{-\tilde g_3} \,, 
\ee
and the uplift term takes the simple form
\be
\label{VbD2}
V^{\overline{D2}}_\text{uplift} = \overline{\mu}_{\overline{D2}} \ e^{-\frac{47}{3}\beta\upsilon-\frac{3}{4}\phi_0} \,, 
\ee
where $\overline{\mu}_{\overline{D2}} \sim (2 \pi)^5 N_{\overline{D2}} M^{-2} s^{4/3}$. The contribution \eqref{VbD2} is directly added to the scalar potential \eqref{Vsusy} to get the total potential. We verified numerically that this uplift allows for shallow meta-stable dS solutions with moduli fixed near the $g_s$ and $\text{vol}(X_7)$  values given in \eqref{GS} and \eqref{VOL} respectively. The existence of such dS further imposes the condition 
\be 
\label{muf}
\overline{\mu}_{\overline{D2}} \approx 0.03\,h^{\frac{17}{6}} m^{\frac{7}{8}}/ f^{\frac{41}{24}} \,.
\ee 
This assumption implies that the size of the uplift terms becomes smaller and smaller at large $f$. This means that we cannot take $f$ as large as we want since we have to match its value from the actual redshifted tension $\overline{\mu}_{\overline{D2}} \sim (2 \pi)^5 N_{\overline{D2}} M^{-2} s^{4/3}$, where we remind the reader that $M$, in contrast to $f$, is bounded by tadpoles. This nicely puts a bound on $f$, which can still be quite large, but blocks the parametric control. The vacuum energy density and the mass matrix eigenvalues are then calculated to be 
\begin{align}
\label{Vf}
    V&\approx 0.8\times 10^{-9}\frac{h^6m^4}{f^8}\,,\\
    V^{''}&\approx \left(10^7,2.4\times 10^5\right)\times V>0 \,.
\end{align}
Clearly, by infinitesimally changing the $0.03$ in \eqref{muf} one changes the numerical prefactor that appears in the dS vacuum energy \eqref{Vf}.  

\section{5. \NoCaseChange{dSillusions}} 

Since our background has fluxes that carry D2 charge, there is an instability towards brane-flux annihilation \cite{Kachru:2002gs}. Remarkably such brane flux annihilations can occur at the perturbative level and to prevent this we require $ N_{\overline{D2}}/M$ to be small enough, of the orders of a few percentages \cite{Gautason:2015tla, Klebanov:2010qs}, a requirement that can overshoot tadpole bounds \cite{Bena:2018fqc, Bena:2020xrh}. Let us verify this now. 

The main obstacle to parametric control is asking that the $A$-cycle volume \eqref{CGLP-tip} is large in string units (for SUGRA control), while $g_s$ remains small. This means that $M$ has to be big enough, yet $KM$ contributes to the O2-tadpole. To make this concrete, note that from equation \eqref{CGLP-tip} we can derive the radius of the $A$-cycle $R_A$, in \emph{string frame}, to be proportional to
\begin{equation}\label{largeM}
R_A \sim M^{3/8}g_s^{11/32}\,.    
\end{equation}
Hence, at weak coupling we need parametric high $M$ while $KM$ contributes to the O2 tadpole constraint. Furthermore, the warping of the throat is set by the exponential in \eqref{s-eq} and so we require
\begin{equation}\label{largeK}
K> Mg_s^{3/4} \sim R_A^{8/3}g_s^{-1/6}\,, 
\end{equation}
This in turn means that $K\gg 1$ and so $MK$ gets parametrically large. Not only is this problematic for the O2 tadpole, but there is a much worse problem: the Romans mass needed for the classical AdS$_3$ vacua induces a local O6 tadpole for every 3-cycle filled with $H_3$ flux from integrating
\begin{equation}
dF_2 = mH_3 + \delta_{O6/D6}\,.    
\end{equation}
Unless we want SUSY-breaking D6 branes in high numbers to cancel the tadpole, we can only hope that a $Z_2$-involution exists allowing a single O6 plane to cancel the RR tadpole of the $H_3$-flux through the B-cycle of the throat. But then $K$ is order $1$. \emph{We conclude there is simply no anti-brane uplift with warped throats based on the classical AdS$_3$ vacua.}

Hence, we must contemplate removing the Romans mass and then we arrive at the classical no-scale Minkowski vacua. One then can hope that the quantum corrections lead to a vacuum that is not living at strong coupling through a quantum correction to the superpotential \eqref{P-SUSY}:
\be
\label{P-SUSY2}
P=\frac{1}{8}e^{-11\beta\upsilon-\frac{\phi}{2}}\left(-f e^{\frac{3}{4}\phi}+e^{\beta\upsilon}h\right) + P_{\text{quantum}} \,.
\ee
This could perhaps come from Euclidean D-brane instantons, or the analogue of gaugino condensation but on the 3d gauge theories living on the bulk D6 branes, if any. However, it is more likely that the leading corrections are perturbative instead since, unlike the 4d situation, there is no holomorphicity to rule out perturbative corrections to $P$. We will assume that corrections can lead to weakly coupled (SUSY) AdS$_3$ vacua and investigate the uplift.  Then, the main tension is combining a small uplift, with a supergravity regime for the tip of the throat, with a non-parametric tadpole constraint. Again, we need large $M$ both for brane-flux stability and a large enough A-cycle at the tip \eqref{largeM}, and we need a small uplift leading to large $K$ through \eqref{largeK} and even larger $KM$-contribution to the O2 tadpole.  We are unaware of a study of O2 tadpole numbers on compact G$_2$ spaces, but one can hope that the tadpoles can be larger than for CY-compactifications since the dimension of the compactification manifold is larger and the holonomy restrictions are less.  

Imagine that one satisfies these rather difficult-looking tadpole conditions, then one still has to make sure that the backreaction of the anti-D2 brane \cite{Giecold:2011gw} does not create a too large local $H_3$-flux density, since otherwise perturbative brane-flux decay still happens \cite{Danielsson:2014yga}. Currently, there is good evidence to show that lethal backreaction is absent \cite{Michel:2014lva, Cohen-Maldonado:2015ssa,Cohen-Maldonado:2016cjh, Blaback:2019ucp, Armas:2018rsy}.  
Goldstino condensation is also another threatening effect for uplifts which however remains much less understood \cite{Farakos:2022jcl}.  

Finally, there is an issue of consistently gluing the throat in the compact G$_2$ space, as was observed for warped CY$_3$ reductions of IIB \cite{Carta:2019rhx, Gao:2020xqh, Lust:2022xoq}. This was first thought to be a problem of making sure that the throat volume, set by local fluxes $K,M$, is smaller than the total volume fixed by the stabilization of global moduli \cite{Carta:2019rhx}. This so-named \emph{throat fitting}-problem turned out to be too naive and the real issue is the warp factor $e^A$ becoming negative in the bulk, leading to unphysical singularities away from the admissible orientifold singularities, known as \emph{the singular bulk}-problem \cite{Gao:2020xqh}. The bounds of the fitting problem are, however, reproduced in the singular bulk problem and so were morally correct. In our case at hand, it is reasonable to expect the same problem.

\section{6. \NoCaseChange{Discussion}}

In an earlier paper \cite{Farakos:2020idt} we observed that the classical AdS$_3$ solutions with scale separation and moduli stabilisation of \cite{Farakos:2020phe} are potentially promising stepping stones for uplifting to de Sitter vacua because there are no tachyons in the AdS$_3$ vacua. The main obstacles in concrete models was found to be the lack of fine-tuning the amount of uplift energy and it was suggested that warped throats should solve that problem. In here, we found that indeed warped throats can solve the fine-tuning, but then lead to unacceptable O6 tadpole problems.

There are two potential ways out of this dSillusion: 1) Either we stay in the framework of IIA on warped G$_2$-spaces with fluxes but remove the Romans mass and its associated O6 tadpole condition. Then one finds classical ``no-scale'' Minkowski solutions with flat directions. Assuming the flat directions are lifted by quantum corrections \eqref{P-SUSY2} and lead to sufficiently weakly coupled AdS$_3$-vacua, we arrive at a 3d analogue to KKLT \cite{Kachru:2003aw} or LVS \cite{Balasubramanian:2005zx}. We then argued that, nonetheless, equations \eqref{largeM}, \eqref{largeK} are unavoidable, and parametric control over the uplift means parametric large O2 tadpole numbers. It is unclear whether that is possible since the O2 tadpoles are topological.  It is interesting to contemplate whether we can give up on the constraint of a large $A$-cycle. This means that the throat tip is out of supergravity control. But since this is a localised problem it is not clear it destabilises the whole set-up. Perhaps there is a string theory computation to verify what happens in that limit, or one relies on the holographic dual set-up \cite{Kachru:2002gs, Klebanov:2010qs} to understand this limit.
2) In case we are not relying on warped throats and go back to ``large SUSY breaking'' with branes in the bulk (for 4d compactifications this has been proposed in \cite{Kallosh:2018nrk,Cribiori:2019bfx}), there was a potential issue with the dependence of the bound flux-stability bound $N_{\overline{D2}}/M\ll 1$ with respect to bulk moduli. In \cite{Farakos:2020idt} we erroneously concluded that this dependence worsens the bound at weak coupling and large volume, whereas it is the other way around; since the decay involves a spherical brane that needs to climb over a cycle, a larger bulk volume means a larger cycle and more energy cost. Similarly, at weak coupling brane tensions are higher and so it ensures perturbative stability.

Regardless of the potential to achieve meta-stable dS$_3$ vacua, we consider our analysis of the embedding of local CGLP throats into compact G$_2$ spaces to be interesting in its own right. 
\\

\begin{acknowledgments}
\emph{ We are very grateful to Cassandra Van der Sijpt for spotting the typo in the stability computation of \cite{Farakos:2020idt} in her master thesis \cite{Cassandra}. GT is supported in part by the NSF grant PHY-2210271 and the Lehigh University CORE grant with grant ID COREAWD40. TVR acknowledges support from FWO Odysseus grant G0F9516N.}
\end{acknowledgments}


\begin{thebibliography}{100}

\bibitem{Coudarchet:2023mfs}
T.~Coudarchet, ``{Hiding the extra dimensions: A review on scale separation in
  string theory},'' \href{http://dx.doi.org/10.1016/j.physrep.2024.02.003}{{\em
  Phys. Rept.} {\bfseries 1064} (2024) 1--28},
  \href{http://arxiv.org/abs/2311.12105}{{\ttfamily arXiv:2311.12105
  [hep-th]}}.

\bibitem{Tringas:2025uyg}
G.~Tringas and T.~Wrase, ``{Scale separation from O-planes},''
  \href{http://arxiv.org/abs/2504.15436}{{\ttfamily arXiv:2504.15436
  [hep-th]}}.

\bibitem{Andriot:2025cyi}
D.~Andriot, N.~Cribiori, and T.~Van~Riet, ``{Scale separation, rolling
  solutions and entropy bounds},''
  \href{http://arxiv.org/abs/2504.08634}{{\ttfamily arXiv:2504.08634
  [hep-th]}}.

\bibitem{Danielsson:2018ztv}
U.~H. Danielsson and T.~Van~Riet, ``{What if string theory has no de Sitter
  vacua?},'' \href{http://dx.doi.org/10.1142/S0218271818300070}{{\em Int. J.
  Mod. Phys. D} {\bfseries 27} no.~12, (2018) 1830007},
  \href{http://arxiv.org/abs/1804.01120}{{\ttfamily arXiv:1804.01120
  [hep-th]}}.

\bibitem{Cicoli:2018kdo}
M.~Cicoli, S.~De~Alwis, A.~Maharana, F.~Muia, and F.~Quevedo, ``{De Sitter vs
  Quintessence in String Theory},''
  \href{http://dx.doi.org/10.1002/prop.201800079}{{\em Fortsch. Phys.}
  {\bfseries 67} no.~1-2, (2019) 1800079},
  \href{http://arxiv.org/abs/1808.08967}{{\ttfamily arXiv:1808.08967
  [hep-th]}}.

\bibitem{VanRiet:2023pnx}
T.~Van~Riet and G.~Zoccarato, ``{Beginners lectures on flux compactifications
  and related Swampland topics},''
  \href{http://dx.doi.org/10.1016/j.physrep.2023.11.003}{{\em Phys. Rept.}
  {\bfseries 1049} (2024) 1--51},
  \href{http://arxiv.org/abs/2305.01722}{{\ttfamily arXiv:2305.01722
  [hep-th]}}.

\bibitem{Junghans:2018gdb}
D.~Junghans, ``{Weakly Coupled de Sitter Vacua with Fluxes and the
  Swampland},'' \href{http://dx.doi.org/10.1007/JHEP03(2019)150}{{\em JHEP}
  {\bfseries 03} (2019) 150}, \href{http://arxiv.org/abs/1811.06990}{{\ttfamily
  arXiv:1811.06990 [hep-th]}}.

\bibitem{Banlaki:2018ayh}
A.~Banlaki, A.~Chowdhury, C.~Roupec, and T.~Wrase, ``{Scaling limits of dS
  vacua and the swampland},''
  \href{http://dx.doi.org/10.1007/JHEP03(2019)065}{{\em JHEP} {\bfseries 03}
  (2019) 065}, \href{http://arxiv.org/abs/1811.07880}{{\ttfamily
  arXiv:1811.07880 [hep-th]}}.

\bibitem{Ooguri:2018wrx}
H.~Ooguri, E.~Palti, G.~Shiu, and C.~Vafa, ``{Distance and de Sitter
  Conjectures on the Swampland},''
  \href{http://dx.doi.org/10.1016/j.physletb.2018.11.018}{{\em Phys. Lett. B}
  {\bfseries 788} (2019) 180--184},
  \href{http://arxiv.org/abs/1810.05506}{{\ttfamily arXiv:1810.05506
  [hep-th]}}.

\bibitem{Hebecker:2018vxz}
A.~Hebecker and T.~Wrase, ``{The Asymptotic dS Swampland Conjecture - a
  Simplified Derivation and a Potential Loophole},''
  \href{http://dx.doi.org/10.1002/prop.201800097}{{\em Fortsch. Phys.}
  {\bfseries 67} no.~1-2, (2019) 1800097},
  \href{http://arxiv.org/abs/1810.08182}{{\ttfamily arXiv:1810.08182
  [hep-th]}}.

\bibitem{Danielsson:2012et}
U.~H. Danielsson, G.~Shiu, T.~Van~Riet, and T.~Wrase, ``{A note on obstinate
  tachyons in classical dS solutions},''
  \href{http://dx.doi.org/10.1007/JHEP03(2013)138}{{\em JHEP} {\bfseries 03}
  (2013) 138}, \href{http://arxiv.org/abs/1212.5178}{{\ttfamily arXiv:1212.5178
  [hep-th]}}.

\bibitem{Andriot:2021rdy}
D.~Andriot, ``{Tachyonic de Sitter Solutions of 10d Type II Supergravities},''
  \href{http://dx.doi.org/10.1002/prop.202100063}{{\em Fortsch. Phys.}
  {\bfseries 69} no.~7, (2021) 2100063},
  \href{http://arxiv.org/abs/2101.06251}{{\ttfamily arXiv:2101.06251
  [hep-th]}}.

\bibitem{Vafa:2005ui}
C.~Vafa, ``{The String landscape and the swampland},''
  \href{http://arxiv.org/abs/hep-th/0509212}{{\ttfamily arXiv:hep-th/0509212}}.

\bibitem{Palti:2019pca}
E.~Palti, ``{The Swampland: Introduction and Review},''
  \href{http://dx.doi.org/10.1002/prop.201900037}{{\em Fortsch. Phys.}
  {\bfseries 67} no.~6, (2019) 1900037},
  \href{http://arxiv.org/abs/1903.06239}{{\ttfamily arXiv:1903.06239
  [hep-th]}}.

\bibitem{Blaback:2010sj}
J.~Blaback, U.~H. Danielsson, D.~Junghans, T.~Van~Riet, T.~Wrase, and
  M.~Zagermann, ``{Smeared versus localised sources in flux
  compactifications},'' \href{http://dx.doi.org/10.1007/JHEP12(2010)043}{{\em
  JHEP} {\bfseries 12} (2010) 043},
  \href{http://arxiv.org/abs/1009.1877}{{\ttfamily arXiv:1009.1877 [hep-th]}}.

\bibitem{Farakos:2020phe}
F.~Farakos, G.~Tringas, and T.~Van~Riet, ``{No-scale and scale-separated flux
  vacua from IIA on G2 orientifolds},''
  \href{http://dx.doi.org/10.1140/epjc/s10052-020-8247-5}{{\em Eur. Phys. J. C}
  {\bfseries 80} no.~7, (2020) 659},
  \href{http://arxiv.org/abs/2005.05246}{{\ttfamily arXiv:2005.05246
  [hep-th]}}.

\bibitem{VanHemelryck:2022ynr}
V.~Van~Hemelryck, ``{Scale-Separated AdS3 Vacua from G2-Orientifolds Using
  Bispinors},'' \href{http://dx.doi.org/10.1002/prop.202200128}{{\em Fortsch.
  Phys.} {\bfseries 70} no.~12, (2022) 2200128},
  \href{http://arxiv.org/abs/2207.14311}{{\ttfamily arXiv:2207.14311
  [hep-th]}}.

\bibitem{Emelin:2021gzx}
M.~Emelin, F.~Farakos, and G.~Tringas, ``{Three-dimensional flux vacua from IIB
  on co-calibrated G2 orientifolds},''
  \href{http://dx.doi.org/10.1140/epjc/s10052-021-09261-y}{{\em Eur. Phys. J.
  C} {\bfseries 81} no.~5, (2021) 456},
  \href{http://arxiv.org/abs/2103.03282}{{\ttfamily arXiv:2103.03282
  [hep-th]}}.

\bibitem{Farakos:2023nms}
F.~Farakos, M.~Morittu, and G.~Tringas, ``{On/off scale separation},''
  \href{http://dx.doi.org/10.1007/JHEP10(2023)067}{{\em JHEP} {\bfseries 10}
  (2023) 067}, \href{http://arxiv.org/abs/2304.14372}{{\ttfamily
  arXiv:2304.14372 [hep-th]}}.

\bibitem{Arboleya:2024vnp}
A.~Arboleya, A.~Guarino, and M.~Morittu, ``{Type II orientifold flux vacua in
  3D},'' \href{http://dx.doi.org/10.1007/JHEP12(2024)087}{{\em JHEP} {\bfseries
  12} (2024) 087}, \href{http://arxiv.org/abs/2408.01403}{{\ttfamily
  arXiv:2408.01403 [hep-th]}}.

\bibitem{VanHemelryck:2025qok}
V.~Van~Hemelryck, ``{Supersymmetric scale-separated AdS$_3$ vacua of type
  IIB},'' \href{http://arxiv.org/abs/2502.04791}{{\ttfamily arXiv:2502.04791
  [hep-th]}}.

\bibitem{Farakos:2025bwf}
F.~Farakos and G.~Tringas, ``{Integer dual dimensions in scale-separated
  AdS$_3$ from massive IIA},''
  \href{http://arxiv.org/abs/2502.08215}{{\ttfamily arXiv:2502.08215
  [hep-th]}}.

\bibitem{DeWolfe:2005uu}
O.~DeWolfe, A.~Giryavets, S.~Kachru, and W.~Taylor, ``{Type IIA moduli
  stabilization},'' \href{http://dx.doi.org/10.1088/1126-6708/2005/07/066}{{\em
  JHEP} {\bfseries 07} (2005) 066},
  \href{http://arxiv.org/abs/hep-th/0505160}{{\ttfamily arXiv:hep-th/0505160}}.

\bibitem{Apers:2022tfm}
F.~Apers, J.~P. Conlon, S.~Ning, and F.~Revello, ``{Integer conformal
  dimensions for type IIa flux vacua},''
  \href{http://dx.doi.org/10.1103/PhysRevD.105.106029}{{\em Phys. Rev. D}
  {\bfseries 105} no.~10, (2022) 106029},
  \href{http://arxiv.org/abs/2202.09330}{{\ttfamily arXiv:2202.09330
  [hep-th]}}.

\bibitem{Dong:2010pm}
X.~Dong, B.~Horn, E.~Silverstein, and G.~Torroba, ``{Micromanaging de Sitter
  holography},'' \href{http://dx.doi.org/10.1088/0264-9381/27/24/245020}{{\em
  Class. Quant. Grav.} {\bfseries 27} (2010) 245020},
  \href{http://arxiv.org/abs/1005.5403}{{\ttfamily arXiv:1005.5403 [hep-th]}}.

\bibitem{Farakos:2020idt}
F.~Farakos, G.~Tringas, and T.~Van~Riet, ``{Classical de Sitter solutions in
  three dimensions without tachyons?},''
  \href{http://dx.doi.org/10.1140/epjc/s10052-020-08525-3}{{\em Eur. Phys. J.
  C} {\bfseries 80} no.~10, (2020) 947},
  \href{http://arxiv.org/abs/2007.12084}{{\ttfamily arXiv:2007.12084
  [hep-th]}}.

\bibitem{Cassandra}
C.~Van~der Sypt, ``{Master thesis at KU Leuven},''.

\bibitem{Kachru:2003aw}
S.~Kachru, R.~Kallosh, A.~D. Linde, and S.~P. Trivedi, ``{De Sitter vacua in
  string theory},'' \href{http://dx.doi.org/10.1103/PhysRevD.68.046005}{{\em
  Phys. Rev. D} {\bfseries 68} (2003) 046005},
  \href{http://arxiv.org/abs/hep-th/0301240}{{\ttfamily arXiv:hep-th/0301240}}.

\bibitem{Carta:2019rhx}
F.~Carta, J.~Moritz, and A.~Westphal, ``{Gaugino condensation and small uplifts
  in KKLT},'' \href{http://dx.doi.org/10.1007/JHEP08(2019)141}{{\em JHEP}
  {\bfseries 08} (2019) 141}, \href{http://arxiv.org/abs/1902.01412}{{\ttfamily
  arXiv:1902.01412 [hep-th]}}.

\bibitem{Gao:2020xqh}
X.~Gao, A.~Hebecker, and D.~Junghans, ``{Control issues of KKLT},''
  \href{http://dx.doi.org/10.1002/prop.202000089}{{\em Fortsch. Phys.}
  {\bfseries 68} (2020) 2000089},
  \href{http://arxiv.org/abs/2009.03914}{{\ttfamily arXiv:2009.03914
  [hep-th]}}.

\bibitem{Giddings:2001yu}
S.~B. Giddings, S.~Kachru, and J.~Polchinski, ``{Hierarchies from fluxes in
  string compactifications},''
  \href{http://dx.doi.org/10.1103/PhysRevD.66.106006}{{\em Phys. Rev. D}
  {\bfseries 66} (2002) 106006},
  \href{http://arxiv.org/abs/hep-th/0105097}{{\ttfamily arXiv:hep-th/0105097}}.

\bibitem{Klebanov:2000hb}
I.~R. Klebanov and M.~J. Strassler, ``{Supergravity and a confining gauge
  theory: Duality cascades and chi SB resolution of naked singularities},''
  \href{http://dx.doi.org/10.1088/1126-6708/2000/08/052}{{\em JHEP} {\bfseries
  08} (2000) 052}, \href{http://arxiv.org/abs/hep-th/0007191}{{\ttfamily
  arXiv:hep-th/0007191}}.

\bibitem{Kachru:2002gs}
S.~Kachru, J.~Pearson, and H.~L. Verlinde, ``{Brane / flux annihilation and the
  string dual of a nonsupersymmetric field theory},''
  \href{http://dx.doi.org/10.1088/1126-6708/2002/06/021}{{\em JHEP} {\bfseries
  06} (2002) 021}, \href{http://arxiv.org/abs/hep-th/0112197}{{\ttfamily
  arXiv:hep-th/0112197}}.

\bibitem{Cvetic:2001bw}
M.~Cvetic, G.~W. Gibbons, J.~T. Liu, H.~Lu, and C.~N. Pope, ``{A New fractional
  D2-brane, G(2) holonomy and T duality},''
  \href{http://dx.doi.org/10.1088/0264-9381/19/20/310}{{\em Class. Quant.
  Grav.} {\bfseries 19} (2002) 5163--5172},
  \href{http://arxiv.org/abs/hep-th/0106162}{{\ttfamily arXiv:hep-th/0106162}}.

\bibitem{Cvetic:2001ih}
M.~Cvetic, G.~W. Gibbons, H.~Lu, and C.~N. Pope, ``{M theory conifolds},''
  \href{http://dx.doi.org/10.1103/PhysRevLett.88.121602}{{\em Phys. Rev. Lett.}
  {\bfseries 88} (2002) 121602},
  \href{http://arxiv.org/abs/hep-th/0112098}{{\ttfamily arXiv:hep-th/0112098}}.

\bibitem{Cvetic:2001kp}
M.~Cvetic, G.~W. Gibbons, H.~Lu, and C.~N. Pope, ``{A G(2) unification of the
  deformed and resolved conifolds},''
  \href{http://dx.doi.org/10.1016/S0370-2693(02)01654-4}{{\em Phys. Lett. B}
  {\bfseries 534} (2002) 172--180},
  \href{http://arxiv.org/abs/hep-th/0112138}{{\ttfamily arXiv:hep-th/0112138}}.

\bibitem{Acharya:2020vmg}
B.~S. Acharya, L.~Foscolo, M.~Najjar, and E.~E. Svanes, ``{New
  G$_{2}$-conifolds in M-theory and their field theory interpretation},''
  \href{http://dx.doi.org/10.1007/JHEP05(2021)250}{{\em JHEP} {\bfseries 05}
  (2021) 250}, \href{http://arxiv.org/abs/2011.06998}{{\ttfamily
  arXiv:2011.06998 [hep-th]}}.

\bibitem{Harvey:1999as}
J.~A. Harvey and G.~W. Moore, ``{Superpotentials and membrane instantons},''
  \href{http://arxiv.org/abs/hep-th/9907026}{{\ttfamily arXiv:hep-th/9907026}}.

\bibitem{Kachru:2001je}
S.~Kachru and J.~McGreevy, ``{M theory on manifolds of G(2) holonomy and type
  IIA orientifolds},''
  \href{http://dx.doi.org/10.1088/1126-6708/2001/06/027}{{\em JHEP} {\bfseries
  06} (2001) 027}, \href{http://arxiv.org/abs/hep-th/0103223}{{\ttfamily
  arXiv:hep-th/0103223}}.

\bibitem{Cvetic:2001ma}
M.~Cvetic, G.~W. Gibbons, H.~Lu, and C.~N. Pope, ``{Supersymmetric nonsingular
  fractional D-2 branes and NS NS 2 branes},''
  \href{http://dx.doi.org/10.1016/S0550-3213(01)00236-X}{{\em Nucl. Phys. B}
  {\bfseries 606} (2001) 18--44},
  \href{http://arxiv.org/abs/hep-th/0101096}{{\ttfamily arXiv:hep-th/0101096}}.

\bibitem{Herzog:2002ss}
C.~P. Herzog, ``{String tensions and three-dimensional confining gauge
  theories},'' \href{http://dx.doi.org/10.1103/PhysRevD.66.065009}{{\em Phys.
  Rev. D} {\bfseries 66} (2002) 065009},
  \href{http://arxiv.org/abs/hep-th/0205064}{{\ttfamily arXiv:hep-th/0205064}}.

\bibitem{Gautason:2015tla}
F.~F. Gautason, B.~Truijen, and T.~Van~Riet, ``{The many faces of brane-flux
  annihilation},'' \href{http://dx.doi.org/10.1007/JHEP10(2015)152}{{\em JHEP}
  {\bfseries 10} (2015) 152}, \href{http://arxiv.org/abs/1505.00159}{{\ttfamily
  arXiv:1505.00159 [hep-th]}}.

\bibitem{Klebanov:2010qs}
I.~R. Klebanov and S.~S. Pufu, ``{M-Branes and Metastable States},''
  \href{http://dx.doi.org/10.1007/JHEP08(2011)035}{{\em JHEP} {\bfseries 08}
  (2011) 035}, \href{http://arxiv.org/abs/1006.3587}{{\ttfamily arXiv:1006.3587
  [hep-th]}}.

\bibitem{Bena:2018fqc}
I.~Bena, E.~Dudas, M.~Gra\~na, and S.~L\"ust, ``{Uplifting Runaways},''
  \href{http://dx.doi.org/10.1002/prop.201800100}{{\em Fortsch. Phys.}
  {\bfseries 67} no.~1-2, (2019) 1800100},
  \href{http://arxiv.org/abs/1809.06861}{{\ttfamily arXiv:1809.06861
  [hep-th]}}.

\bibitem{Bena:2020xrh}
I.~Bena, J.~Bl\r{a}b\"ack, M.~Gra\~na, and S.~L\"ust, ``{The tadpole
  problem},'' \href{http://dx.doi.org/10.1007/JHEP11(2021)223}{{\em JHEP}
  {\bfseries 11} (2021) 223}, \href{http://arxiv.org/abs/2010.10519}{{\ttfamily
  arXiv:2010.10519 [hep-th]}}.

\bibitem{Giecold:2011gw}
G.~Giecold, E.~Goi, and F.~Orsi, ``{Assessing a candidate IIA dual to
  metastable supersymmetry-breaking},''
  \href{http://dx.doi.org/10.1007/JHEP02(2012)019}{{\em JHEP} {\bfseries 02}
  (2012) 019}, \href{http://arxiv.org/abs/1108.1789}{{\ttfamily arXiv:1108.1789
  [hep-th]}}.

\bibitem{Danielsson:2014yga}
U.~H. Danielsson and T.~Van~Riet, ``{Fatal attraction: more on decaying
  anti-branes},'' \href{http://dx.doi.org/10.1007/JHEP03(2015)087}{{\em JHEP}
  {\bfseries 03} (2015) 087}, \href{http://arxiv.org/abs/1410.8476}{{\ttfamily
  arXiv:1410.8476 [hep-th]}}.

\bibitem{Michel:2014lva}
B.~Michel, E.~Mintun, J.~Polchinski, A.~Puhm, and P.~Saad, ``{Remarks on brane
  and antibrane dynamics},''
  \href{http://dx.doi.org/10.1007/JHEP09(2015)021}{{\em JHEP} {\bfseries 09}
  (2015) 021}, \href{http://arxiv.org/abs/1412.5702}{{\ttfamily arXiv:1412.5702
  [hep-th]}}.

\bibitem{Cohen-Maldonado:2015ssa}
D.~Cohen-Maldonado, J.~Diaz, T.~van Riet, and B.~Vercnocke, ``{Observations on
  fluxes near anti-branes},''
  \href{http://dx.doi.org/10.1007/JHEP01(2016)126}{{\em JHEP} {\bfseries 01}
  (2016) 126}, \href{http://arxiv.org/abs/1507.01022}{{\ttfamily
  arXiv:1507.01022 [hep-th]}}.

\bibitem{Cohen-Maldonado:2016cjh}
D.~Cohen-Maldonado, J.~Diaz, and F.~F. Gautason, ``{Polarised antibranes from
  Smarr relations},'' \href{http://dx.doi.org/10.1007/JHEP05(2016)175}{{\em
  JHEP} {\bfseries 05} (2016) 175},
  \href{http://arxiv.org/abs/1603.05678}{{\ttfamily arXiv:1603.05678
  [hep-th]}}.

\bibitem{Blaback:2019ucp}
J.~Blaback, F.~F. Gautason, A.~Ruiperez, and T.~Van~Riet, ``{Anti-brane
  singularities as red herrings},''
  \href{http://dx.doi.org/10.1007/JHEP12(2019)125}{{\em JHEP} {\bfseries 12}
  (2019) 125}, \href{http://arxiv.org/abs/1907.05295}{{\ttfamily
  arXiv:1907.05295 [hep-th]}}.

\bibitem{Armas:2018rsy}
J.~Armas, N.~Nguyen, V.~Niarchos, N.~A. Obers, and T.~Van~Riet, ``{Meta-stable
  non-extremal anti-branes},''
  \href{http://dx.doi.org/10.1103/PhysRevLett.122.181601}{{\em Phys. Rev.
  Lett.} {\bfseries 122} no.~18, (2019) 181601},
  \href{http://arxiv.org/abs/1812.01067}{{\ttfamily arXiv:1812.01067
  [hep-th]}}.

\bibitem{Farakos:2022jcl}
F.~Farakos and M.~Morittu, ``{Goldstino condensation at large N},''
  \href{http://dx.doi.org/10.1140/epjc/s10052-023-11330-3}{{\em Eur. Phys. J.
  C} {\bfseries 83} no.~2, (2023) 166},
  \href{http://arxiv.org/abs/2211.12527}{{\ttfamily arXiv:2211.12527
  [hep-th]}}.

\bibitem{Lust:2022xoq}
S.~L\"ust and L.~Randall, ``{Effective Theory of Warped Compactifications and
  the Implications for KKLT},''
  \href{http://dx.doi.org/10.1002/prop.202200103}{{\em Fortsch. Phys.}
  {\bfseries 70} no.~7-8, (2022) 2200103},
  \href{http://arxiv.org/abs/2206.04708}{{\ttfamily arXiv:2206.04708
  [hep-th]}}.

\bibitem{Balasubramanian:2005zx}
V.~Balasubramanian, P.~Berglund, J.~P. Conlon, and F.~Quevedo, ``{Systematics
  of moduli stabilisation in Calabi-Yau flux compactifications},''
  \href{http://dx.doi.org/10.1088/1126-6708/2005/03/007}{{\em JHEP} {\bfseries
  03} (2005) 007}, \href{http://arxiv.org/abs/hep-th/0502058}{{\ttfamily
  arXiv:hep-th/0502058}}.

\bibitem{Kallosh:2018nrk}
R.~Kallosh and T.~Wrase, ``{dS Supergravity from 10d},''
  \href{http://dx.doi.org/10.1002/prop.201800071}{{\em Fortsch. Phys.}
  {\bfseries 67} no.~1-2, (2019) 1800071},
  \href{http://arxiv.org/abs/1808.09427}{{\ttfamily arXiv:1808.09427
  [hep-th]}}.

\bibitem{Cribiori:2019bfx}
N.~Cribiori, R.~Kallosh, C.~Roupec, and T.~Wrase, ``{Uplifting
  Anti-D6-brane},'' \href{http://dx.doi.org/10.1007/JHEP12(2019)171}{{\em JHEP}
  {\bfseries 12} (2019) 171}, \href{http://arxiv.org/abs/1909.08629}{{\ttfamily
  arXiv:1909.08629 [hep-th]}}.


\end{thebibliography}

\end{document}